\documentstyle[12pt]{article}
\begin{document}

\baselineskip 12pt

\begin{center}
{\bf EXCITATION SCATTERING IN INTEGRABLE MODELS \\
AND HALL-LITTLEWOOD-KEROV POLYNOMIALS} \\

\bigskip
Peter G. O. Freund \\
{\it Enrico Fermi Institute and Department of Physics \\
University of Chicago, Chicago, IL 60637} \\

\bigskip

and \\

\bigskip

Anton V. Zabrodin \\
{\it Institute of Chemical Physics \\
Kosygina Str. 4, SU-117334, Moscow, Russia} \\

\end{center}

\bigskip
\noindent
ABSTRACT\\
The S-matrices for the scattering of two excitations in the XYZ model and
in all of its $SU(n)$\--type generalizations are obtained from the asymptotic
behavior
of Kerov's generalized Hall\--Littlewood polynomials.
These physical scattering processes are all reduced to geometric
$s$\--wave scattering problems on certain quantum\--symmetric spaces,
whose zonal spherical functions these Hall\--Littlewood\--Kerov polynomials
are. Mathematically, this involves a generalization with an unlimited number
of parameters of the Macdonald polynomials. Physically, our results suggest
that, of the $(1+1)$\--dimensional models, the integrable ones are those,
for which the scattering of excitations becomes geometric in the sense above.
\bigskip
\medskip

S-matrices in (1+1)-dimensional integrable models have been extensively
studied for quite some time now [1]-[5]. Recently [6,7],we proposed
a new geometric approach to this study. Avoiding (like references [4,5])
complicated Bethe ansatz calculations and dealing directly with physical states
in the thermodynamic limit, we showed that the scattering
of dressed excitations in a vast class of integrable models is equivalent
to the $s$-wave scattering of a particle on certain, in general {\it quantum}
(i.e. non-commutative) symmetric spaces. Namely, the Jost functions for the
corresponding input four-point amplitudes (two-excitation scattering) in
integrable spin chains have been identified [6,7] with the explicitly
calculable Harish-Chandra c-functions of those symmetric spaces.
This observation
is in good agreement with ideas of Connes [8] about the role of
non-commutative geometry in physics: the quantum mechanics of a particle on a
{\it non-commutative} space provides us with some information about quantum
field theory.

   The hamiltonian of a free particle on a classical (rank 1) symmetric space
 coincides
with the group invariant Laplace-Beltrami operator on that space. In order to
solve the stationary scattering problem, one must find the eigenfunctions of
the Laplace-Beltrami operator corresponding to the continuous spectrum, and
then find their asymptotic behavior at spatial infinity. For $s$-wave
scattering,we must consider only the spherically symmetric (i.e. K-invariant
from the left, in the
general case of a rank 1 non-compact coset space G/K) eigenfunctions of the
"radial part" of the Laplace-Beltrami operator acting on the double coset
space $K\backslash G/K$. Such functions are called zonal spherical functions
(zsf's) on
G/K and their asymptotic behavior at large values of the radial variable
is determined by the Harish-Chandra c-function [9].

   Remarkably, this picture can be extended to the case of quantum symmetric
spaces, the analog of the Laplace-Beltrami operator being a difference
operator.
The harmonic analysis on quantum groups was developed in references [10-12].
The
zsf's on the "quantum hyperboloid" $H_{q}=[SL(2,R)/SO(2)]_{q}$ are given [11],
[6] by
the properly normalized continuous q-Legendre polynomials which can be
expressed through the q-hypergeometric series [13]. These polynomials are
orthogonal with respect to a certain continuous measure on the unit circle and
satisfy a three-term recurrence relation [13]. This relation can be interpreted
as the eigenvalue equation for the quantum analog of the radial part of the
Laplace-Beltrami operator [6,7]. The corresponding Harish-Chandra function is
 expressed in terms of the q-gamma function [13] and gives the physical
S-matrix in the XXZ spin chain.

   In this paper we develop this approach further and extend it to the most
general integrable spin-1/2 chain, the XYZ model with elliptic R-matrix and
its $SU(n)$ generalizations [3]. It is well-known that this model is
equivalent to the famous 8-vertex model on the square lattice, solved by
Baxter [14]. The excitations in the XYZ
spin chain were studied in reference [15]. Later the model was also treated
with the quantum inverse scattering method [16]. The S-matrix for the
scattering of two physical excitations in the XYZ model (actually its
eigenvalue corresponding to the properly symmetrized state with two holes in
the physical vacuum)
can be calculated by means of Korepin's method [2] (see also the second
article in ref. [1]). It can be represented in the form
$$
                   S(u)=J(iu)/J(-iu)
\eqno(1)
$$
with the Jost function
$$
J(iu)=\prod_{k=0}^\infty

\frac{\Gamma_{q}(iu+rk)\Gamma_{q}(iu+rk+r+1)}{\Gamma_{q}(iu+rk+1/2)\Gamma_{q}(iu+rk+r+1/2)}
\eqno(2)
$$
where
$$
\Gamma_{q}(x)=(1-q)^{1-x}\prod_{k=0}^\infty
\frac{(1-q^{k+1})}{(1-q^{k+x})}
\eqno(3)
$$
is the q-deformed gamma-function [13] which tends to the ordinary gamma
function as $q \rightarrow 1$. In (2) we used the following notation:
$$
q=\exp (-4\gamma ),~~~~ r=- \frac{i\pi\tau}{2\gamma}
\eqno(4)
$$
where $\gamma$ and $\tau$ are the parameters of the elliptic
 R-matrix,the
anisotropy and elliptic module respectively, and $u$ is the properly rescaled
spectral parameter which is interpreted as the relative rapidity of the
particles. Our notations here differ from those of references [7]!
Specifically,
our $q$ was called $t^2$ and our $r$ was called $(2l)^{-1}$ there. Equation (2)
above is obtained (modulo one irrelevant "q-Blaschke" factor) directly from
equation
(6b) of the first paper [7]. The existence of this alternate form of the
equation is closely related to the symmetry displayed in equation (4) of
that paper.
   Note that in the limit $\tau\rightarrow i\infty$ (the case of the
antiferromagnetic XXZ
model with trigonometric R-matrix) we reproduce from (2) the Harish-Chandra
function of the quantum hyperboloid $H_{q}$ [6]:
$$
    c(iu)=\frac{\Gamma_{q}(iu)}{\Gamma_{q}(iu+1/2)}
\eqno(5)
$$
The further limit $q\rightarrow 1$ gives the well-known classical result [9].
 If we first
take the limit $q\rightarrow 1$, keeping r fixed, we would obtain essentially
the
soliton-soliton S-matrix in the sine\--Gordon model [1,2] (a proper
 redefinition
of parameters is implied).

   The intriguing question is, in what symmetric space does the function
$J(iu)$ of equation (2) play the role of Harish\--Chandra c-function.
 Clearly, it is some 2-parameter quantum deformation of the
classical hyperboloid $H=SL(2,R)/SO(2)$. We are going to construct a family of
orthogonal polynomials $P_{n}$ which are further deformations of the
continuous
q-Legendre polynomials and can be considered as zsf's on that hypothetical
"symmetric space". As a confirmation of this, we shall show that the
corresponding "Harish\--Chandra c\--function" (explicitly calculable from the
asymptotic behaviour of $P_{n}$ at large $n$) is just the Jost function
$J(iu)$ of equation (2).

A natural framework for this construction is the theory of
Hall\--Littlewood\--Kerov (HLK) polynomials [17]. They provide a natural
infinite\--parameter deformation of Schur functions and include many of
the known zsf's
on classical (real, as well as $p$\--adic) and quantum symmetric spaces as
special, or limiting cases. In particular, Macdonald's polynomials [18] for
root systems $A_n$ are all special HLK polynomials.

   We now briefly describe the HLK polynomials [17]. We use the
 notations from Macdonald's book [19]. Let A be the algebra of
 symmetric polynomials in infinitely many variables
$x=(x_{1}, x_{2},...)$. One convenient linear basis in A is the set of all
products of power sums
$$
p_{\lambda}(x)=\prod_{{i=1}\atop{\lambda_i>0}}{\sum_{j=1}}
(x_{j}^{\lambda_{i}})
\eqno(6)
$$
where $\lambda$ denotes a Young diagram, $\lambda_{i}$ being the length of its
i-th row so that $\lambda_i=0$ for sufficiently large i. The product in (6)
runs
over those i with $\lambda_{i}>0$, so it is always finite. There are also other
important bases in A (elementary symmetric functions $e_{\lambda}$, monomial
functions $m_{\lambda}$, Schur functions $s_{\lambda}$, etc...) with the
property
that the basis vectors are indexed by Young diagrams. In the case of $n$
 non\--zero variables $x_{1},\ldots, x_{n}$, the Schur functions can be
defined as follows (for more details see [19]);
$$
s_{\lambda}(x) = \frac{ \det (x_i^{\lambda_j +n-j })}{\det (x_{i}^{n-j})}, ~~~
1 \leq i,j \leq n,
\eqno(7)
$$
On the set $Y(n)$ of Young diagrams with n boxes the following total
order is
defined [17]: for $\lambda , \mu \in Y(n)$, we have $\lambda > \mu$ if the
first non-vanishing difference $\lambda_{i}-\mu_{i}$ is positive.

Fix a sequence of formal parameters (complex numbers)
$w=(w_{1}, w_{2},\ldots)$ and put
$$
w_{\lambda}=\prod_j w_j^{m_{j}(\lambda)}
\eqno(8)
$$
$$
z_{\lambda}= \prod_j j^{m_{j}(\lambda)} (m_{j}(\lambda))!
\eqno(9)
$$
where $m_{j}(\lambda )$ is the number of rows of length $j$ in the Young
diagram $\lambda$.

Introduce a scalar product $< , >_w$ on A by
$$
<p_{\lambda},~ p_{\mu} >_{w} = \delta_{\lambda ,\mu} z_{\mu} w_{\mu}
\eqno(10)
$$
with $p_\lambda$ given by eq. (6).

Now we are in a position to define the HLK polynomials $P_{\lambda}(x;w),~
\lambda \in Y(n)$, as the result of the orthogonalization of the basis
formed by the Schur functions $s_\lambda$with respect to the scalar
product (10) and the
total order $<$ :
$$
P_{\lambda}(x;w) = s_{\lambda}(x)+ \sum_{\mu <\lambda} a_{\lambda ,\mu}(w)
s_{\mu}(x)
\eqno(11)
$$
$$
<P_{\lambda}, ~P_{\mu} >_w=0~~~~~~ {\rm if}~~ \lambda \neq \mu
\eqno(12)
$$
(in particular $P_{\lambda} (x,1) =s_{\lambda}(x)$).
In this sense,
the HLK polynomials can be considered as {\it deformations of Schur
functions}. The
deformation parameters are $w_{1},~ w_{2},\ldots$ (in general, all of them are
independent).

Interesting applications arise when we parametrize the whole infinite set
of $w_{j}$'s in terms of a {\it finite} number of new parameters. Let us give
an important example. Set
$$
w_{k}= \frac{1-q^k}{1-t^k}, ~~~~~~~    k=1,2, \ldots
\eqno(13)
$$
with the parameters $q,t$ obeying $|q|<1,~ |t| <1$ ($q=0$ corresponds to the
classical Hall-Littlewood case). Then in the case
of $n$ non\--zero variables $x_{1},\ldots ,x_{n}$ with the constraint
$x_{1} x_{2} \ldots x_{n}= 1$, the $ P_\lambda$'s reduce to Macdonald
polynomials for root system
$A_{n-1}$ [18]. Their ``geometrical'' meaning is that of zsf's on
various --- in general quantum- --- symmetric spaces.

The most interesting feature of all this is, that we can ``interpolate''
between different types of symmetric spaces varying the two parameters
$q,t$.
In particular, the limit $t=q^{l} \rightarrow 1$ ($l$ fixed) gives, for
certain values of $l$, the zsf's on real symmetric spaces.
For $q=0$, $t=1/p$ ($p=$ prime) these $P_{\lambda}$'s
essentially reduce to the zsf's on $p$\--adic symmetric spaces.
In [7] we argued that the values ``in between'' these two cases
yield zsf's on some
{\it quantum} symmetric spaces (in general related to a certain limit of
Sklyanin-Odesskii-Feigin
algebras [20], [21]).
Physically, the scattering problems on these spaces describe
the scattering of dressed excitations in generalized anisotropic
$SU(n)$\--magnetics, associated with the $Z_n$ Baxter model.

Keeping in mind the connection with root systems, we will refer to the case of
$n$ variables as the $GL(n)$\--like case.
It is then sufficient to consider Young diagrams with no more than
$n$ non\--empty rows.

Let $b_{\lambda} =b_{\lambda}(w)$ be the inverse square of the norm of
$P_{\lambda}$:
$$
b_{\lambda}= \frac{1}{<P_{\lambda}, ~P_{\lambda}>_{w}}.
\eqno(14)
$$
For $\lambda=(n)$, the one\--row Young diagram of length $n$, call
$P_{n}(x;w)$ and $b_n$ the corresponding HLK polynomial and its inverse
square norm (equation (14)) respectively. Then
$$
    W(z)=\exp (\sum_{n=1}^\infty \frac {z^{n}}{nw_{n}})
\eqno(15)
$$
generates the polynomials $b_{n}P_{n}$, since [17]
$$
\sum_n b_{n}P_{n}(x;w)z^{n} = \prod_i W(x_{i} z)
\eqno(16)
$$
As we shall see, $W(z)$ plays a very important role in the theory and is
closely related to the ``generalized Harish\--Chandra function'', which we
seek. One can show that
$$
W(z)= \sum_{ n=0}^\infty b_{n}z^{n}.
\eqno(17)
$$

Define the ``structure constants'' $f_{\mu, \nu}^{\lambda}=f_{\nu
,\mu}^{\lambda}$:
$$
P_{\mu}P_{\nu}=\sum_\lambda  f_{\mu, \nu}^{\lambda} P_{\lambda}.
\eqno(18)
$$
We shall make use of the following \\
{\it Theorem} [17]: for $f_{\mu, \nu}^{\lambda}\neq 0$,
 we have $\mu \nu \leq \lambda \leq  \mu +\nu$
(the diagrams $\mu \nu$ and ($\mu + \nu$) consist of the rows (${\mu_{i},
\nu_{j}}$) and
(${\mu_{i}+\nu_{i}}$), respectively).\\
We specialize to the $GL(2)$\--like case
which leads to orthogonal polynomials in {\it one} variable.
In fact, it is possible to obtain more or less explicit representations of
HLK polynomials
for two\--row Young diagrams.
Physically, this case correspond to {\it two} colliding
``particles'' (excitations).

We would like to emphasize that there are two, a priori {\it different}, though
probably isomorphic, $GL(2)$\--deformations in the problem. One of them
determines
the structure of the $R$\--matrix and $L$\--operator. For the XYZ model
it is the Sklyanin algebra [20], a two-parameter deformation of GL(2).
The other one is
a (quantum) group\--like object with ``zsf's'' given by special HLK
polynomials (to be
specified below) and with the Harish\--Chandra function of the form (2). It
appears that we deal with two faces, algebraic and geometric, of one and
the same object, as is the case for the XXZ spin chain [6].

In the rest of the paper we study the following natural generalization of
the choice (13), which turns out to be connected with the XYZ model:
$$
w_k = \frac{(1-q_1^k)(1-q_2^k)}{(1-t_1^k)(1-t_2^k)},  ~~~      k=1,2,\ldots
\eqno(19)
$$
with $|q_{i}|<1, |t_{i}|<1$. Most of the results are valid for more
general $w_{k}$ given by a {\it finite} product of ratios of the same kind
with a
finite number of $(q_{i}, t_{i})$ pairs. However, the general case is much
more
complicated technically, so from now on we restrict ourselves
to $w_k$ of the form  (19).

For two-row
Young diagrams, it is convenient to use instead of the row lengths
$\lambda_1 , \lambda_2$ the two quantities $D$ and $n$, related to
$\lambda_i$ by $ \lambda_{1}=D+n,~~ \lambda_{2}=D$.
We shall call $D$ the {\it depth} of the Young diagram.
All the quantities carrying index
$\lambda$ will
be denoted as $P_{n}^{(D)}, b_{n}^{(D)}$, etc... The non\--zero variables are
$x_{1}=z, x_{2}=z^{-1}$. More precisely, we use the following prescription:
$x_{1}=\rho z, x_{2}=\rho z^{-1}$ with $|z|=1, |\rho |<1$ and $\rho \rightarrow
1$ at the end of the calculation.
Clearly, $P_{n}^{(D)}(z,z^{-1}) (=:P_{n}^{(D)}(z)$ for brevity) is a
Laurent polynomial in $z$ of degree $n$, normalized so, that
$P_{n}^{(D)}(z)=z^{n}+z^{-n}+ \ldots$
For $z=\exp (i \theta )$, $P_{n}^{(D)}$ is a
polynomial of degree $n$ in
$x=\cos \theta $

A remarkable fact, proved in [17,18], is that for $w_{k}$ given by (13)
(Macdonald case), the $P_{n}^{(D)}$'s {\it do not depend} on $D$
(the same is true for all the limiting cases of (13)).
Then one can prove that
$P_{n}^{(D)}(z)=:P_{n}(z)$
are orthogonal on the unit circle (or on the segment [-1,1] in terms of the
variable x) and coincide with the Rogers-Askey-Ismail (RAI) polynomials
[22].
The continuous $q$\--Legendre functions are a special case of them.

The case (19) is more complicated, because now, the polynomials do in general
depend on $D$ and are no longer orthogonal for finite $D$.
We are therefore going to construct {\it orthogonal} polynomials
from the whole family ${P_{n}^{(D)}}$
(the orthogonality is necessary for the interpretation of our polynomials as
zsf's!).

To this end, we use the relation (18) and Kerov's theorem mentioned after
it, specialized to the case at hand.
Notice first that for all $D,~~ P_{0}=1,~~ P_{1}=2x=z+ z^{-1}$. Then
$$
(z+ z^{-1})P_{n}^{(D)}=P_{n+1}^{(D)}+c_{n-1}^{(D+1)} P_{n-1}^{(D+1)}
                                           ~~~  ~ ~  n=1,2, \ldots
\eqno(20)
$$
with some coefficients $c_{k}^{(D)}$. This resembles a 3-term recurrence
relation, but there is the shift of the depth in the last term. If there were
no dependence on $D$ we would get a true recurrence relation and the
orthogonality would then follow. How could we manufacture orthogonal
polynomials starting from (20)?

The hint is to consider the limit $D \rightarrow \infty$. Then, at least
formally, the limiting polynomials
$$
P_{n}(z)= \lim_{ D \rightarrow \infty}  P_{n}^{(D)}(z)
\eqno(21)
$$
satisfy a 3-term recurrence relation and hence are orthogonal.

One can prove that the limit (21) exists and
that the $P_{n}$'s, which we shall
call {\it Generalized Macdonald} (GM) polynomials, are well defined.
The details will be given elsewhere [23]. Here we only show how to determine
the corresponding
measure in the orthogonality relation (an analog of the Plancherel measure).

In order to do this let us return for a moment to the RAI polynomials and
recall
the main idea of how they are identified with the zsf's on a quantum
hyperboloid.
The RAI polynomials $C_{n}(\cos \theta )$ possess the following self\--duality
property
[7]: $C_{n}(\cos(\theta ))$ are zsf's on the (non\--compact) $q$\--hyperboloid
when $n$ is considered as a coordinate variable and $\theta$ as the spectral
parameter; at
the same time they are zsf's on the compact Cartan dual space $S_{q}=
(SU(2)/SO(2))_{q}$ (quantum sphere [11]) when $\theta$ is a coordinate and $n$
is the spectral index.
So dealing with GM polynomials $P_{n}(\cos \theta )$ it is
natural
to consider $n$ as a coordinate variable on some more general non\--compact
quantum space.

There is a standard method in harmonic analysis for obtaining the Plancherel
measure $d \mu$ from known zsf's [8].
The Plancherel measure is completely
determined by the asymptotic behaviour of zsf's and is expressed through
the Harish\--Chandra $c$\-- function as follows: $d{\mu}(u)=du/|c(u)|^{2}$.
Similar arguments work in our situation.

The large $n$ asymptotics of the GM polynomials (21) can be found from (16).
Note that the {\it leading} terms in the asymptotics of $P_{n}^{(D)}$ do
{\it not}
depend on $D$
(this is clear from (20)), so it is enough to find the asymptotics of
$P_{n}^{(0)}$ by contour integrating both sides of (16) and
then extracting the leading terms.
The result is
$$
 P_{n}(z)|_{ n \rightarrow \infty} =z^{n}W(z^{-2})+z^{-n}W(z^{2})+~
 {\rm exponentially~ small~ terms}
\eqno(22)
$$
with the $W$-function given by eq. (15).
This looks like a superposition of incoming and outgoing waves.
{}From (22) we can read off the ``Harish\---Chandra $c$\--function''.
By definition it is
$$
                   C(z)=W(z^{2})
\eqno(23)
$$
We set (see equations (4))
$$
q_{1}=q,~    q_{2}=q^{r},~    t_{1}=q^{1/2},~  t_{2}=q^{r+1/2}
\eqno(24)
$$
in equation (19). With the so obtained $w_k$, we then calculate  $W(z)$,
according to equation (15), and find from equation (23) that the function
$C(z)$ is identical with the Jost function $J(iu)$ of equation (2) for
$z=q^{iu/2}$. {\it This realization of the $S$\--matrix of the XYZ model
is our main result.} This result admits  considerable generalization. By
changing equation (24) to
$$
q_{1}=q,~    q_{2}=q^{r},~   t_{1}=q^{1/n},~   t_{2}=q^{r+1-{1/n}},
\eqno(25)
$$
with integer $n\ge 2$, $q=\exp (-2n\gamma)$ and $r=-i\pi\tau/n\gamma$,
we can calculate a new $C(z)$ function, in the same
way. This function then yields the Jost function for the scattering of two
lowest level excitations in the generalized anisotropic SU($n$) magnetics,
associated to the $Z_n$ Baxter model, the case $n=2$ being that
of the XYZ model treated above. In the more general case (25), the four
parameters $q_{1}, q_{2}, t_{1}, t_{2}$ are expressed in terms of the three
parameters $q, r, n$, rather than the two parameters $q, r$, which appear
in equation (24). Even for the more general case (25), one therefore
has one relation between the four parmeters, to wit
$$
q_{1}q_{2}=t_{1}t_{2}.
\eqno(26)
$$
The polynomials $P_{n}(z)$ obey a three-term recurrence relation obtained
from (20) in the limit (21):
$$
  P_{n+1}(z)+c_{n-1}^{\infty } P_{n-1}(z)=(z+z^{-1})P_{n}(z)
\eqno(27)
$$
This might be interpreted as the eigenvalue equation (with the eigenvalue
$z+z^{-1}$) for the ``radial part'' of some difference Laplace\--Beltrami
operator (the
l.h.s.).
Using standard arguments we can conclude from (22) and (27) that the $P_{n}$'s
are orthogonal with respect to the Plancherel\--like measure
$$
     d \mu (z)= \frac{(W(z^{2})W(z^{-2}))^{-1}}{(4\pi iz)} dz
\eqno(28)
$$
on the unit circle:
$$
<<P_{n},P_{m}>>:= \int d \mu(z)P_{n}(z)P_{m}(z^{-1})=\delta_{mn} h_{n}
\eqno(29)
$$
The computation of the square norm $h_{n}$ is a very interesting unsolved
problem.

Finally, let us point out the connection between the
axiomatically defined Kerov scalar
product $<,>_w$ (10)  and the scalar product
$<<,>>$
(29) given explicitly as a contour integral. The connection we
found is
$$
<<P_{n}(z),P_{n}(z)>>=b_{\infty}^{2} \lim_{ D \rightarrow \infty}
                      <P_{n}^{(D)},P_{n}^{(D)}>_{w}
\eqno(30)
$$
where $b_{\infty }= \lim_{n \rightarrow \infty}  b_{n}$. This important
relation can be
proved using the determinant representation of the polynomials $P_{n}^{(D)}$
to be
presented in [23].

In conclusion, we would like to make a few remarks.

1. It would be very interesting to reformulate the above theory in terms
of the Toda chain hierarchy, using the well\--known connection between the
formalism
of orthogonal polynomials and the integrable hierarchies, recently studied in
detail
in the context of matrix models [24]. Here are some fragments of such a
connection.
The relation (27) is the Lax equation for the Toda chain, Kerov's
parameters $w_{k}$ are essentially the Toda flows (``times'' $T_{k}): T_{k}=
1/(kw_{k})$, whereas $q_{i}, t_{i}$ in (19) are related to Miwa's
variables, the square norm $h_{n}$ (29) is the ratio of two $\tau$\--functions
$\tau_{n}/\tau_{n-1}$ of the Toda hierarchy, etc...

2. It is very likely that HLK-like polynomials provide a general framework for
understanding the $S$\--matrices in {\it all} the known quantum integrable
models. For generalized anisotropic $SU(n)$\--magnetics in {\it higher}
representations, it is necessary to include more $(q_{i},t_{i})$-like
parameters in (13)
and (19), essentially coresponding to the Casimir invariants which identify
the higher representation. On the other hand, we have shown [23] that
the $S$\--matrix for the scattering of two physical
excitations of different {\it kind} in the anisotropic generalized
$SU(n)$\--magnetics
with Belavin's $n^{2} \times n^{2}$ elliptic $R$\--matrix [25] (in
the fundamental representation)
is given by the above construction with {\it three} pairs of parameters
$(q_{i}, t_{i})$ obeying the obvious generalization of the relation (26).
The two new parameters, which join the three parameters already in place (
equation (25)), specify the levels of the two scattering excitations.

3. From the mathematical point of view, HLK polynomials describe some
deformations of
$GL(2)$ with an arbitrary number of parameters. Although for a general
deformation
we can't yet interpret the corresponding ``symmetric space'' geometrically,
the HLK
functions provide well-defined families of ``zsf's'' on these ``would\--be
spaces''
possessing the necessary formal properties.

In this paper, we considered in some detail such a $GL(2)$\--deformation with
{\it two pairs} of independent parameters $(q_{i}, t_{i})$ and showed how
it is
connected with the XYZ model. On the other hand, such deformations should
be closely related to quantum affine Kac\--Moody algebras studied recently in
[26].

As for further deformations, we can speculate that each time one
deforms not $GL(2)$ itself but some ever more "extended" algebraic objects
containing $GL(2)$ (similarly to the familiar case of the one\--parameter
quantum
deformation, where one deforms not the Lie algebra itself, but its universal
enveloping algebra). Though we do not know how these deformations might
look, it seems remarkable that all the deformation parameters have a
direct physical meaning in integrable systems.

4. The HLK polynomials have allowed us to reduce the problem of excitation
scattering in a vast class of (1+1)\--dimensional models to that of
$s$\--wave-scattering on some quantum-symmetric space. The trivial S-matrix
$S=1$ describes $s$\--wave "scattering" on the simplest symmetric space,
euclidean space.  On other symmetric or quantum\--symmetric spaces,
scattering is less trivial, but still purely {\it geometric}, as the
potential vanishes. This appears to be a generic feature: {\it of the
(1+1)\--dimensional models, the integrable ones are those for which the
scattering of excitations becomes geometric}.

We are grateful to A.Gorsky, D.Lebedev, A.Mironov, A.Olshanetsky,
A.Zelevinsky and especially to A.Berenshtein for discussions, and to the
Mathematical Disciplines Center, University of Chicago for generous support.
This work was supported in part by a grant from the NSF: PHY-91-23780.

\medskip
\noindent
{\bf References}

\small

\bigskip
%1.
\noindent [1] A. B. Zamolodchikov, and
Al. B. Zamolodchikov,  Ann. Phys. (N.Y.) {\bf 120} (1979) 253;
A. B. Zamolodchikov, Comm. Math. Phys. {\bf 69} (1979) 165.
\smallskip

%2
\noindent [2] V. E. Korepin, Theor. Math. Phys. {\bf 41} (1979) 953.
\smallskip

%3
\noindent [3] P. P. Kulish, and N. Yu. Reshetikhin,  Soviet Phys. JETP {\bf
53} (1981) 108.
\smallskip

%4
\noindent [4] D. Bernard, A. LeClair, Comm. Math. Phys {\bf142} (1991) 99
\smallskip

%5
\noindent [5] B. Davies, O. Foda, M. Jimbo, T. Miwa, A. Nakayashiki,
RIMS preprint, 1992.
\smallskip

%6
\noindent [6] A. V. Zabrodin, Mod. Phys. Lett. {\bf A7} (1992) 441.
\smallskip

%7
\noindent [7] P. G. O. Freund, A. V. Zabrodin, Phys. Lett. {\bf 284B} (1992)
283;
Comm. Math. Phys. {\bf147} (1992) 277.
\smallskip

%8
\noindent [8] A. Connes, Pub. Math. IHES {\bf62} (1985) 257.
\smallskip

%9
\noindent [9] S. Helgason,  {\em Topics in Harmonic Analysis on Homogenous
Spaces},
Birkh\-\"{a}user, Basel, 1981.
\smallskip

%10
\noindent [10] L. L. Vaksman, and Ya. S. Soibelman, Funct. Anal. Appl. {\bf
22} (1988)
170; L. L. Vaksman, and L. I. Korogodsky, Funct. Anal. Appl. {\bf 25} (1991)
48.
\smallskip

%11
\noindent [11] T. Koornwinder, Nederl. Akad. Wetensch. Proc. {\bf A92} (1989)
97;
in {\em Orthogonal Polynomials Theory and Practice},
P.~Nevai, ed., Kluwer Academic Publ., Dordrecht, 1990, p. 257.
\smallskip

%12
\noindent [12] T. Masuda, K. Mimachi, Y. Nakagami, M. Noumi and K. Ueno,
C. R. Acad. Sci. Paris, Ser. I Math. {\bf 307} (1988) 559.
\smallskip

%13
\noindent [13] G. Gasper, and M. Rahman,  {\em Basic Hypergeometric Series},
Cambridge
Univ. Press, Cambridge, 1990.
\smallskip

%14
\noindent [14] R. J. Baxter,  {\em Exactly Solved Models in Statistical
Mechanics},
Acad. Press, N.Y., 1982.
\smallskip

%15
\noindent [15] J. Johnson, S. Krinsky, B. McCoy, Phys. Rev. {\bf A8} (1973)
2526.
\smallskip

%16
\noindent [16] L. A. Takhtadjan and L. D. Faddeev, Uspekhi Mat. Nauk, {\bf 34}
(1979) 13.
\smallskip

%17
\noindent [17] S. V. Kerov, Funct. Anal. Appl. {\bf 25} (1991) 65.
\smallskip

%18
\noindent [18] I. G. Macdonald,  in {\em Orthogonal Polynomials:
Theory and Practice}, P.~Nevai ed., Kluwer Academic Publ., Dordrecht, 1990;
Queen Mary College preprint 1989.
\smallskip

%19
\noindent [19] I. G. Macdonald {\em Symmetric Functions and Hall Polynomials},
Clarendon, Oxford, 1979.
\smallskip

%20
\noindent [20] E. K. Sklyanin, Funk. Anal. Appl. {\bf 16} (1982) 263;
Funk. Anal. Appl. {\bf 17} (1983) 273.
\smallskip

%21
\noindent [21] A. V. Odeskii, and B. L. Feigin, Funk. Anal. Appl. {\bf 23}
(1989) 207
\smallskip

%22
\noindent [22] L. J. Rogers, Proc. London Math. Soc. {\bf 26}(1895) 318;
R. Askey, and M. E. -H. Ismail,  in {\em Studies in Pure Math.}
P.~Erd\"{o}s, ed.
Birkh\"{a}user, Basel, 1983.
\smallskip

%23
\noindent [23] P. G. O. Freund, A. V. Zabrodin, in preparation.
\smallskip

%24
\noindent [24] A. Gerasimov, A. Marshakov, A. Mironov, A. Morozov, A. Orlov,
Nucl.
Phys. {\bf B357} (1991) 565; S. Kharchev, A. Marshakov, A. Mironov,
A. Orlov, A. Zabrodin, Nucl. Phys. {\bf B366} (1991) 569.
\smallskip

%25
\noindent [25] A. A. Belavin, Nucl. Phys. {\bf B180} (1981) 189.
\smallskip

%26
\noindent [26] I. Frenkel and N. Yu Reshetikhin, Comm. Math. Phys. {\bf 146}
(1992) 1.

\end{document}